\documentclass[aps,prb,twocolumn,groupedaddress,showpacs]{revtex4}
\usepackage{graphicx}

\input epsf.sty
\begin{document}
\draft

\title{Doping dependence of the vortex glass and sublimation transitions in the high-$T_{c}$ superconductor La$_{2-x}$Sr$_{x}$CuO$_{4}$ as determined from macroscopic measurements}

\author{
        R. Gilardi, S. Streule, and J. Mesot
       }
\affiliation{
         Laboratory for Neutron Scattering; ETH Zurich and PSI 
             Villigen; CH-5232 Villigen PSI; Switzerland}

\author{N. Momono, and M. Oda}
\affiliation{Departement of Physics; Hokkaido University; Sapporo 
         060-0810 Japan}

\date{\today}

\begin{abstract}
Magnetization and ac-susceptibility measurements are used to 
characterize the mixed phase of the high-temperature cuprate 
superconductor La$_{2-x}$Sr$_{x}$CuO$_{4}$ over a large range of doping (0.075~$\leq~x\leq$~0.20). The first order vortex lattice phase transition line $H_{FOT}(T)$, the upper 
critical field $H_{c2}(T)$ and the second peak $H_{sp}(T)$ have been 
investigated up to high magnetic fields (8~Tesla applied perpendicular to the $CuO_2$ planes).
Our results reveal a strong doping dependence of the magnetic phase diagram, which can mainly be 
explained by the increasing anisotropy with underdoping.
Within our interpretation, the first order vortex lattice phase transition is due to the sublimation (rather than melting) of the vortex lattice into a gas of pancake vortices, whereas the second peak is related to the transition to a more disordered vortex glass state.  
\typeout{polish abstract}
\end{abstract}

\maketitle
\narrowtext

\section{Introduction}
Despite belonging to the family of the first high-T$_{c}$ 
superconductor (HTSC) to be discovered, 
the magnetic phase diagram of La$_{2-x}$Sr$_{x}$CuO$_{4}$ (LSCO) has not been as 
intensively investigated as that of other cuprates such as YBa$_{2}$Cu$_{3}$O$_{x}$ (YBCO) and 
Bi$_{2}$Sr$_{2}$CaCu$_{2}$O$_{8+x}$ (BSCCO). 
The LSCO compound has a relative small value of $T_c$ (38.5 K at optimal
doping), but is of high interest because it fills the gap between rather 3D 
systems such as YBCO and highly anisotropic 2D systems such as 
BSCCO. The anisotropy factor $\gamma^{2}$ can be defined as the ratio 
between the out-of-plane and the in-plane resistive components 
($\rho_{c}/\rho_{ab}$) measured in the normal state \cite{SASAGAWA1,SASAGAWA2}. 
An additional advantage of LSCO is that $\gamma$ depends on 
the Sr content $x$ and allows a study of the magnetic phase diagram over a wide 
range of anisotropy (200 $<\gamma^{2}<$ 4000) which lies inbetween the values 
for YBCO (25 $<\gamma^{2}<$ 100) and 
BSCCO (3000 $<\gamma^{2}<$ 30000).
\newline The magnetic phase diagram of HTSC cuprates is 
dominated by the mixed phase (the lower critical field $H_{c1}$(0~K) is 
about 10$^{-2}$ T whereas the upper critical field $H_{c2}$(0~K) is of the 
order of 10$^2$ T), where the magnetic flux can penetrate into the sample 
in the form of quantized flux-lines (vortices).  Due to the anisotropy 
and thermal fluctuations one observes a number of vortex phases, which 
have been the subject of extended experimental and theoretical 
research in the last two decades \cite{BLATTER}.  
In LSCO one can distinguish between a
first order transition (FOT) line $H_{FOT}(T)$, which has been attributed to the melting \cite{BRANDT,BLATTER,HOUGHTON}
or sublimation \cite{SASAGAWA1,SASAGAWA2} of the vortex lattice into a
vortex fluid, and the irreversibility line $H_{irr}(T)$, where reversible
magnetization and resistivity appear \cite{SASAGAWA1,SASAGAWA2, ANDO}.
Another interesting feature is the so-called \textit{fishtail effect}, that is an anomalous second peak in the magnetization loops. The origin of the second peak line $H_{sp}(T)$ is controversial, and has been attributed to mechanisms varying from dimensional 
crossover\cite{TAMEGAI}, 
collective pinning \cite{KRUSIN}, crossover between different pinning phases, 
crossover to a disordered vortex glass \cite{GIAMARCHI,KOSHELEV,GILLER}, etc.
\newline Only recently the vortex lattice (VL) has been directly observed in  overdoped LSCO by means of small angle neutron scattering (SANS), revealing a field-induced transition from 
hexagonal to square symmetry \cite{GILARDI1,GILARDI2} and the vanishing of the VL signal at temperatures well below $T_{c2}$ \cite{GILARDI3}. In the underdoped regime of LSCO, on the other hand, a more disordered vortex glass has been observed by means of muon spin rotation ($\mu$SR) experiments \cite{DIVAKAR}.
Interestingly, recent inelastic neutron scattering (INS) experiments indicate a possible interplay between the vortex and copper-spin degrees of freedom.
In optimally doped LSCO, sub-gap spin excitations induced by a magnetic field of 7.5 Tesla have been
observed at low-temperatures \cite{LAKE01}. Moreover, the spin gap was found to close at the
irreversibility temperature rather than at $T_{c2}$ \cite{LAKE01,GILARDI4}.  In underdoped LSCO, field-induced static incommensurate magnetic peaks have been observed \cite{LAKE02}, and it has been suggested that these field-induced magnetic signals arise from antiferromagnetic order in the vortex cores and in the surrounding regions \cite{AROVAS,DEMLER,HU}. Enhanced antiferromagnetic spin correlations in
the vortex core region have been indeed observed in NMR experiments \cite{MITROVIC,KAKUYANAGI}. 
\newline In order to understand these experiments performed in the presence of an external magnetic field, 
it is crucial to have a good knowledge of the rich and complicated magnetic phase diagram of HTSC.  
We will present here a detailed study of the doping dependence of the magnetic phase diagram in LSCO single crystals from a macroscopic point of view.

\section{Experimental}

The magnetic phase diagram of LSCO has been investigated by means of magnetization ($M$) and ac-susceptibility ($\chi$) measurements.  We used a Quantum Design Physical Properties Measurements System (PPMS) up to fields of 8 T applied approximately perpendicular to the $CuO_2$ planes.  The angle $\Theta$ between the field direction and the c-axis of the samples was always smaller than 10 degrees. This precision is good enough for the present study, since the critical lines (e.g. melting line $H_m$, upper critical field $H_{c2}$) are known to be only slightly affected by small $\Theta$ angles (e.g. $H_{m}(\Theta)\sim H_{m}(\Theta=0)/cos(\Theta)$, $H_{c2}(\Theta)\sim H_{c2}(\Theta=0)/cos(\Theta)$)\cite{Welp,Schilling}. 
\newline Four high quality LSCO single crystals with different doping levels have been 
measured. Details of the sample growth can be found elsewhere\cite{ODA}.
The samples are labeled by the doping region (OD for overdoped and UD for underdoped) together with their $T_{c}$, defined by $\chi'(T_c)=\frac{1}{2}\chi'$(0~K). The width of the superconducting transition $\Delta T_c$ has been determined by the 10\%-90\% criterion. 
OD-31K is a highly overdoped (x=0.20, $T_{c}$=31.5 K, $\Delta T_c$=2.8 K) 51 mg crystal.
OD-36K is slightly overdoped (x=0.17, $T_{c}$=36.2 K, $\Delta T_c$=1.5 K) and is 
a portion of the crystal used for our SANS and INS experiments 
\cite{GILARDI1,GILARDI4}. While most of 
the measurements on the OD-36K sample have been performed on a 293 mg cylindrical 
crystal, for zero-field cooled magnetization measurements the crystal has been cut to a 84 mg plate-like shape with the c-axis parallel to the largest face, in order to reduce the diamagnetic signal. UD-29K is an underdoped (x=0.10, $T_{c}$=29.2 K, $\Delta T_c$=1.3 K) 37~mg plate-like crystal with the c-axis parallel to the largest face, which has been cut from a larger crystal used in $\mu$SR experiments \cite{DIVAKAR}.
Finally, UD-19K is a highly underdoped (x=0.075, $T_{c}$=19 K, $\Delta T_c$=3.8 K) 52 mg 
plate-like crystal with the c-axis perpendicular to the largest face.

\section{Results}
We start with the complex ac-susceptibility $\chi$=$\chi'$+i$\chi''$. The 
samples are placed in an external magnetic field 
$H_{ext}=H_{dc}+H_{ac} \cdot cos(\omega _{ac} t)$, 
with $H_{ac}$=10 Oe and
$\omega _{ac}$~=~10~Hz, 0~T$\leq H_{dc}\leq$8~T.  
\begin{figure}[t]
\includegraphics[width=1.75in]{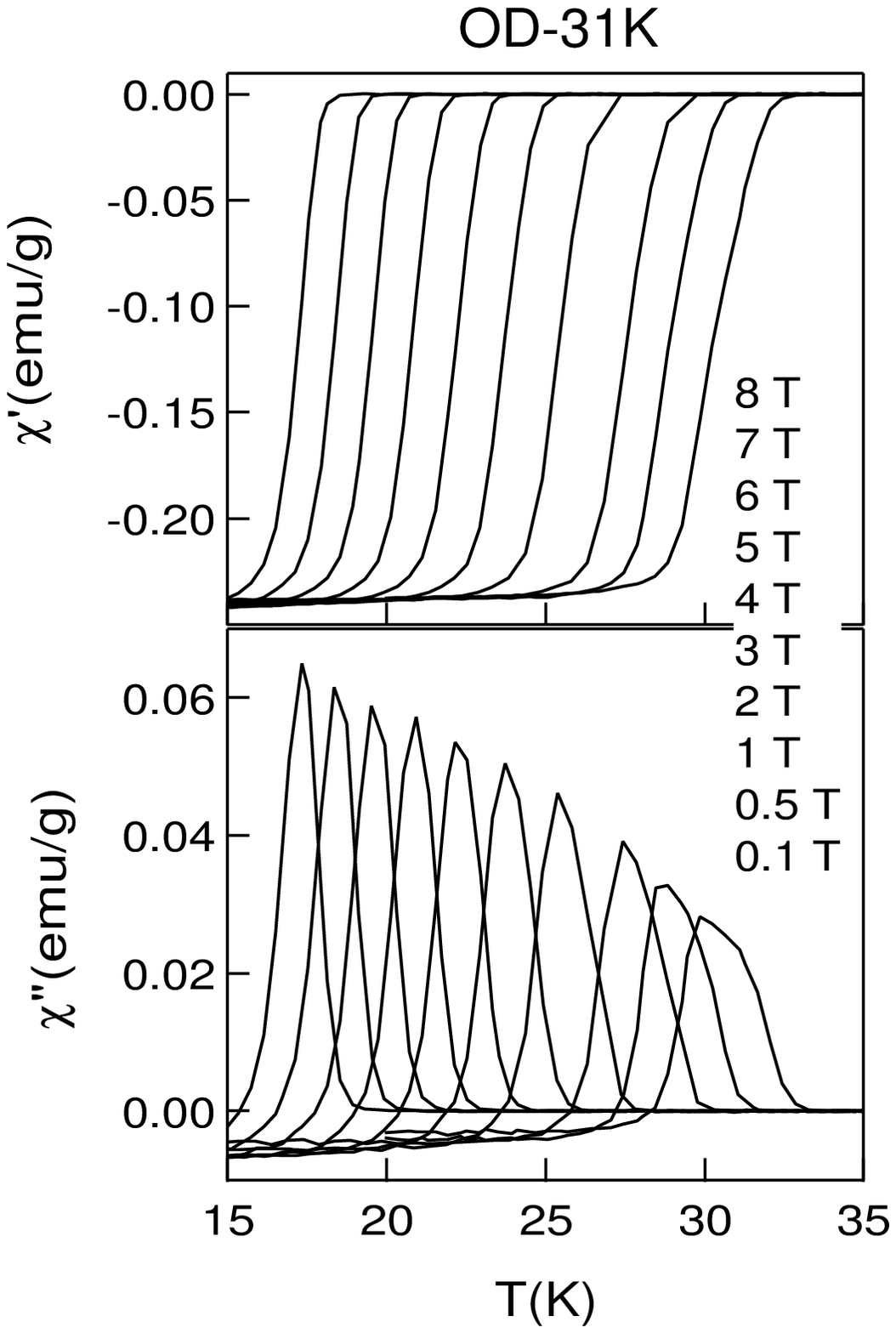}
\hspace{-10mm}
\includegraphics[width=1.9in]{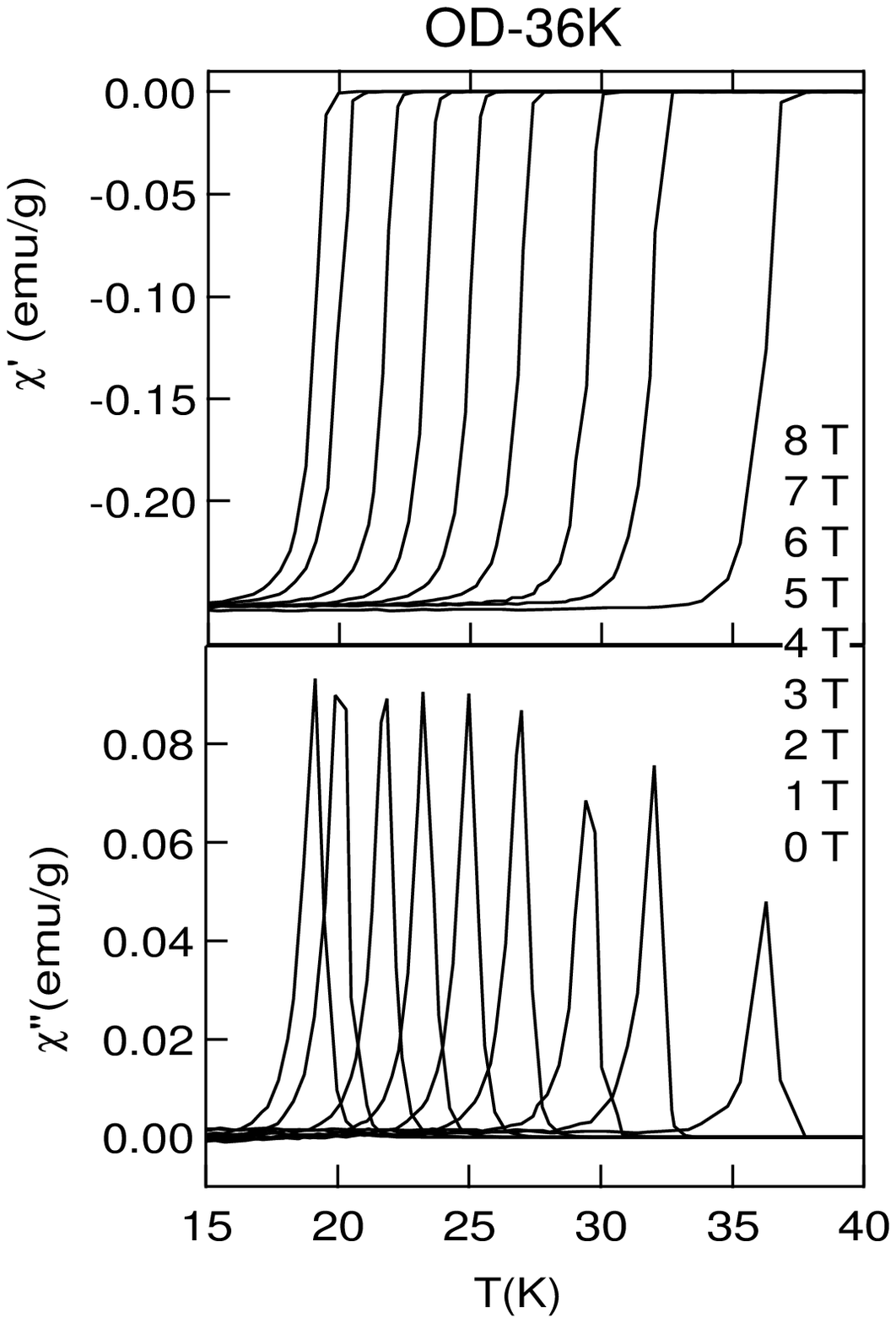}
\includegraphics[width=1.77in]{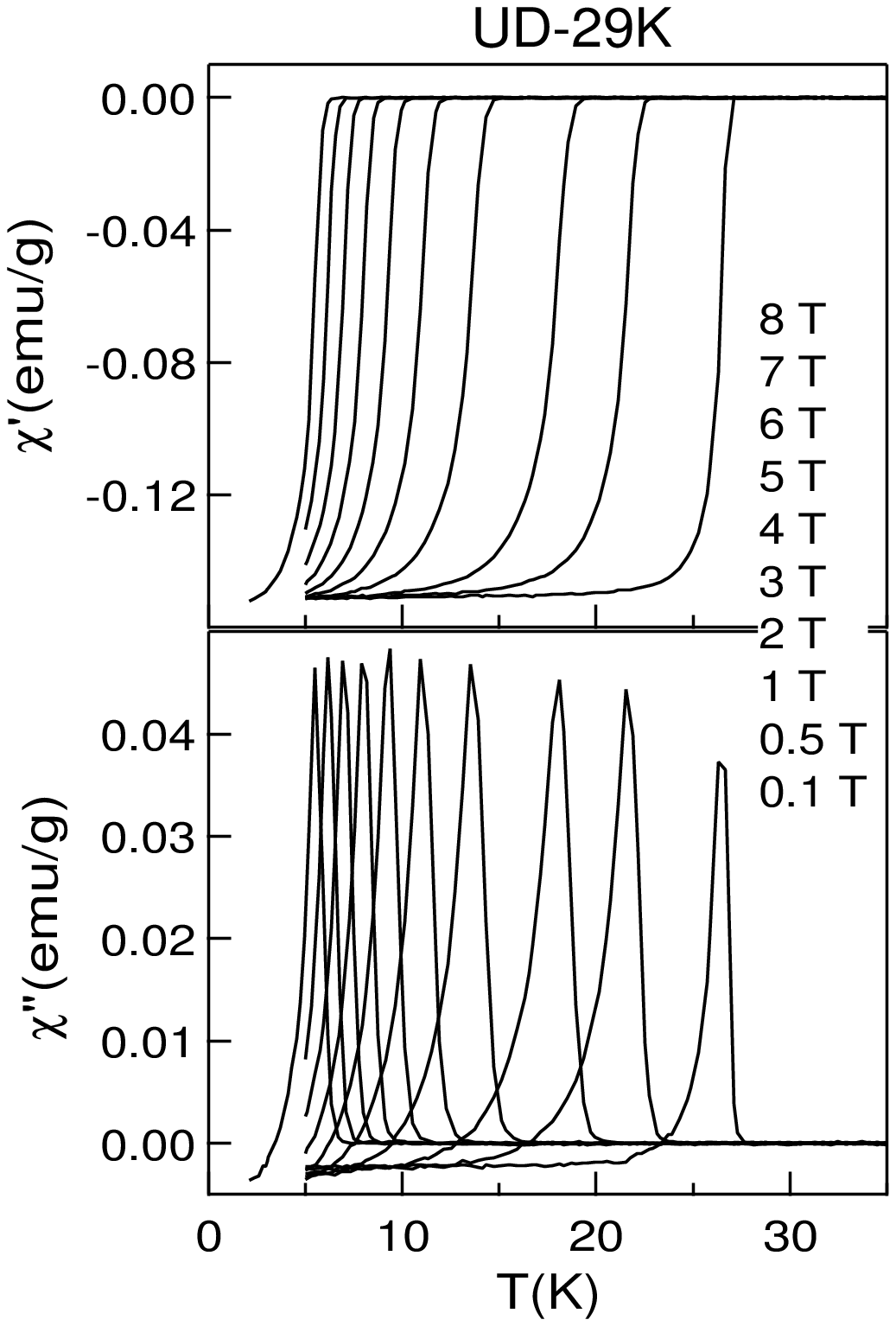}
\hspace{-6mm}
\includegraphics[width=1.73in]{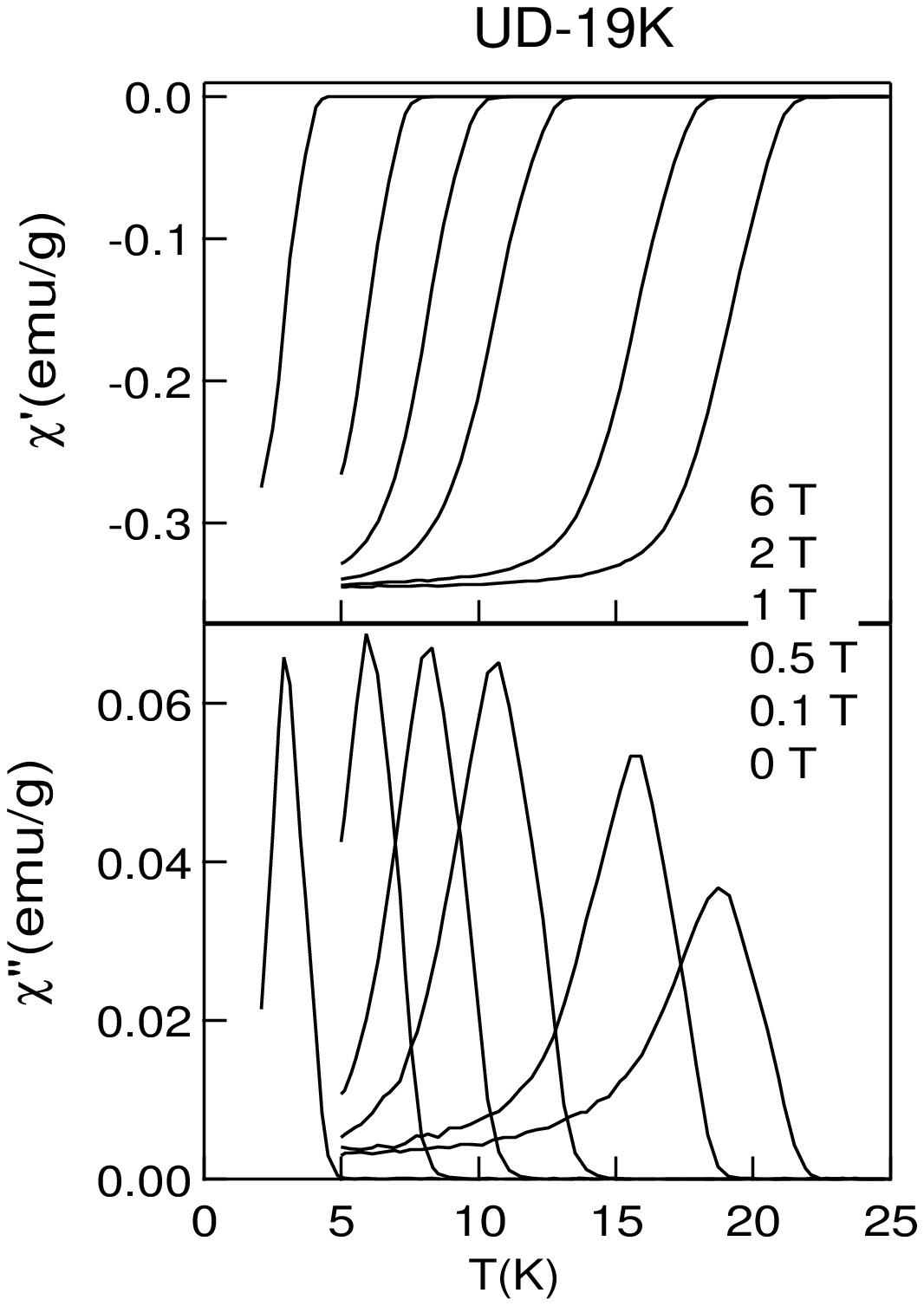}
\caption{ Real and imaginary 
part of the ac-susceptibility $\chi(T)$ for  OD-31K, OD-36K, UD-29K and UD-19K measured at different magnetic fields between 0 T and 8 T. The peak in $\chi''(T)$ rapidly shifts toward lower temperatures with increasing field.}
\label{fig1}
\end{figure}
A set of field-cooled (FC) temperature scans $\chi(T)$ for the four LSCO samples in 
different magnetic fields is shown in Fig.$\ref{fig1}$, with the real part $\chi'$ and imaginary part $\chi''$ plotted separately. 
In all samples the peak in $\chi''$ shifts toward low temperatures and sharpens with increasing magnetic field. However, the magnitude of the shift is strongly doping dependent: for UD-19K  a magnetic field of 6 T is sufficient to shift the peak by 0.85$\cdot T_c$, whereas for OD-31K
the shift caused by a field of 8 T is only 0.45$\cdot T_c$. The detailed field dependence will be
discussed in Section \ref{discussion}.
 In Fig.$\ref{fig2}$ a representative curve $\chi(T)$ measured at $H_{dc}$=3 T for UD-29K is plotted together with magnetization curves $M(T)$. One can notice that there is no difference between the zero-field cooled (ZFC) and the FC $\chi(T)$ data, whereas FC and ZFC $M(T)$ curves separate below the irreversibility temperature $T_{irr}$. Slightly above $T_{irr}$ there is a jump in the magnetization, indicating the presence of a first order transition (FOT) \cite{SASAGAWA1,SASAGAWA2}. Similar data have been obtained for the other samples and for other values of $H_{dc}$.
The jump is more pronounced at high magnetic fields, and in UD-19K only a broad anomaly could be observed (to note is that in this sample the loss peaks in $\chi''(T)$ are very broad, as well).
\newline The experimental $T_{irr}$ is often obtained 
from the loss peak in $\chi''(T)$, which is directly related to the 
maximum slope in $\chi'(T)$ \cite{GOMORY}. 
However, in our case, $T_{irr}$ obtained by ac-susceptibility measurements is slightly higher than the "real" $T_{irr}$, and is concomitant to the jump in $M(T)$ at $T_{FOT}$. 
The irreversibility line and the FOT line are found to be close to each other in all the samples, and are therefore strongly related to each other.  In the following we will consider only the FOT line in the phase diagram.
\begin{figure}
\includegraphics[width=3in]{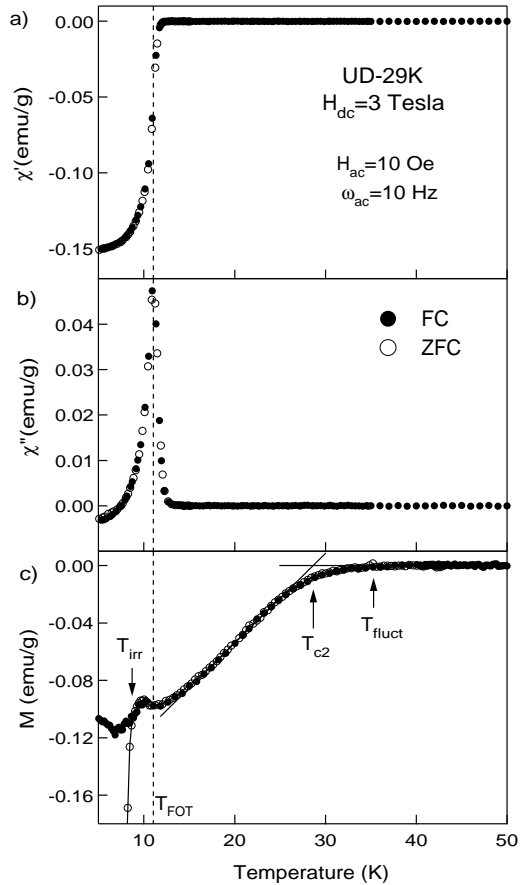}
 \caption{ a) Real and b) imaginary part of the ac-susceptibility 
  $\chi(T)$  of 
 UD-29K in an external field of 3 T.  $T_{FOT}$ is determined by the peak position in $\chi''(T)$, which 
 corresponds to the maximum slope in $\chi'(T)$. c) Magnetization data measured at $H_{dc}$=3 T, after subtraction of a linear background taken in the normal state.   Below $T_{irr}$ the FC and ZFC $M(T)$ curves separate, whereas at $T_{FOT}$ a jump is observed. $T_{c2}$ is estimated by extrapolation (see text). $T_{fluct}$ is defined as the temperature where the data deviates from the horizontal normal state line.}
\label{fig2}
\end{figure}
\begin{figure}
\includegraphics[width=3in]{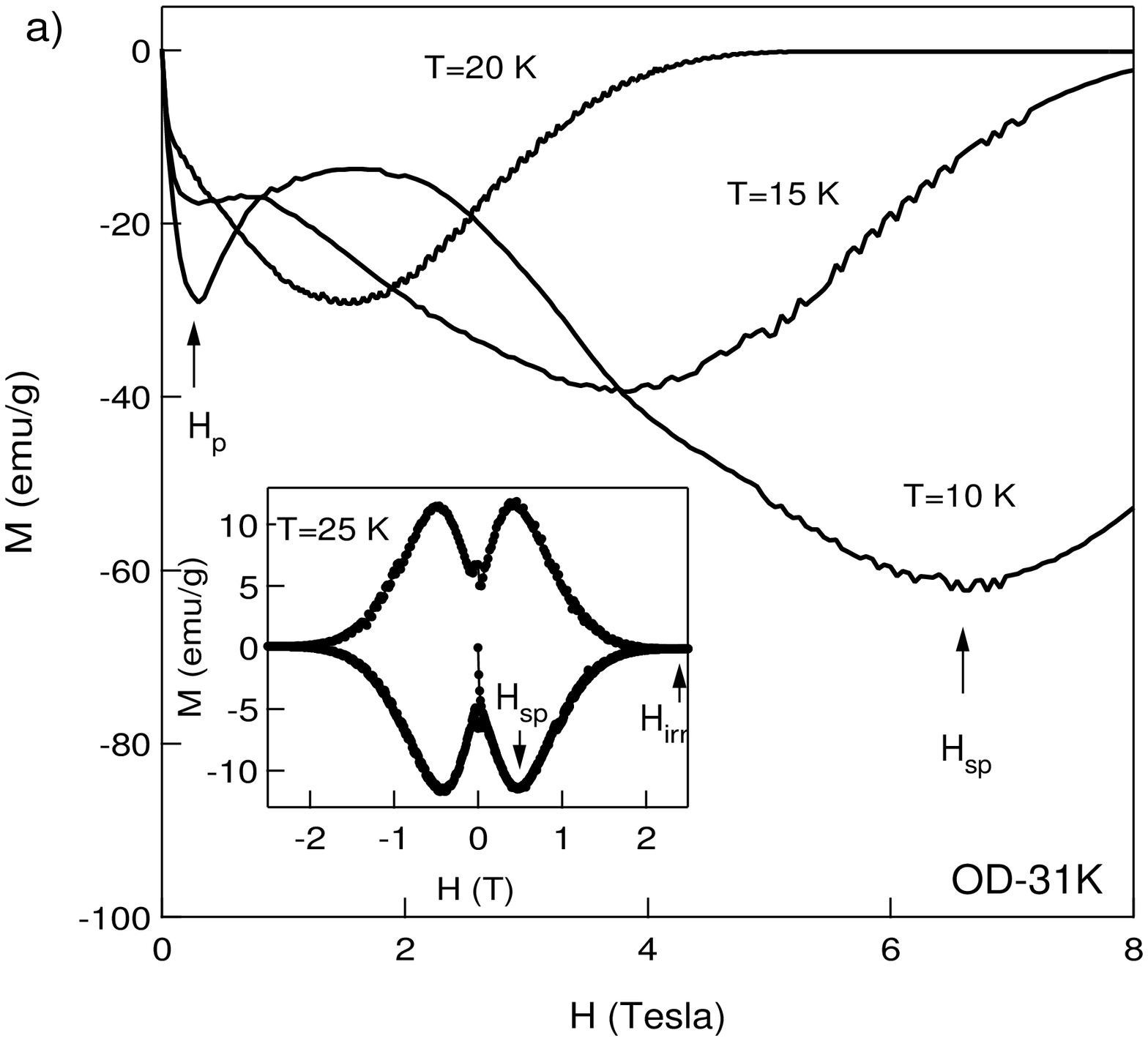}
\includegraphics[width=3in]{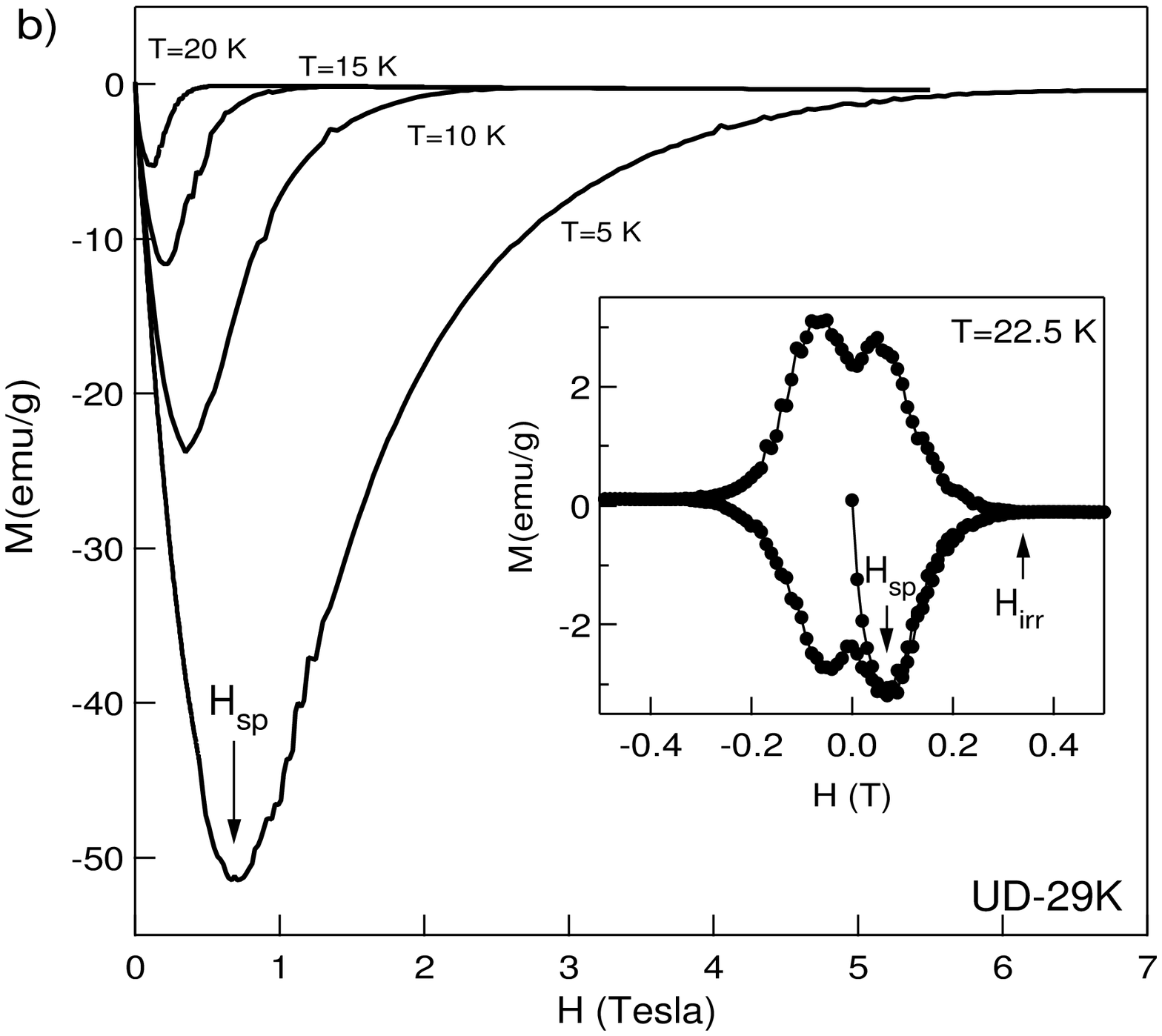}
\caption{ZFC isothermic magnetization curves for a) OD-31K and b) UD-29K. 
For OD-31K, $H_{p}$ and $H_{sp}$ have been determined as 
indicated by the arrows. For UD-29K, only $H_{sp}$ could be observed. 
The insets show full hysteresis loops with $H_{irr}$.}
\label{fig3}
\end{figure}
\begin{figure}
\includegraphics[width=3in]{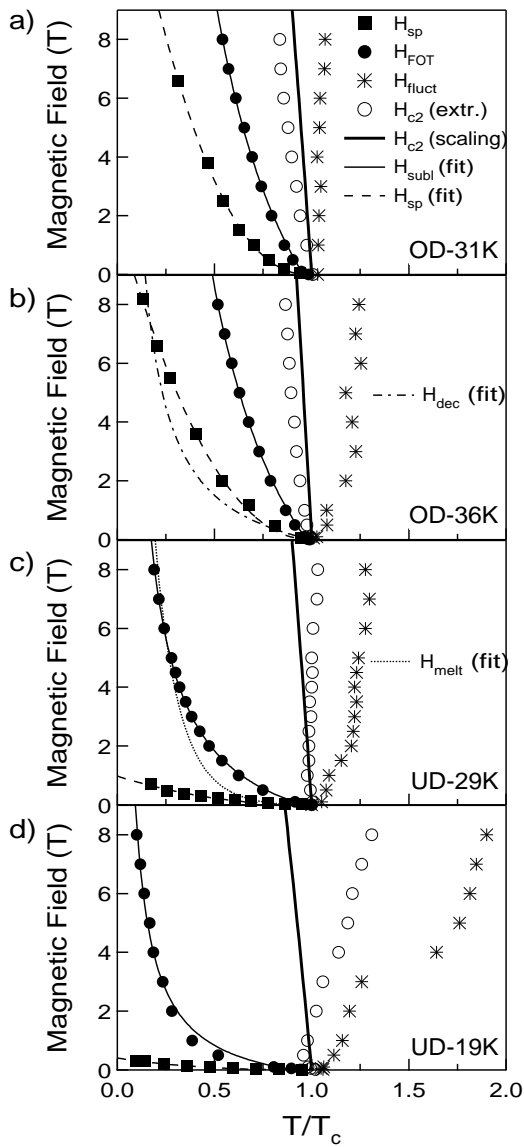}
\caption{Magnetic phase diagram of the four LSCO samples (OD-31K, OD-36K, UD-29K, UD-19K) showing the temperature dependencies of the second peak field $H_{sp}(T)$, the FOT line $H_{FOT}(T)$, the upper critical field $H_{c2}(T)$ (determined by extrapolation and by the scaling procedure), and the field $H_{fluct}(T)$ where diamagnetic fluctuations set in. In a)-d) the second peak line has been fitted by the power law (Eq.(\ref{power})), whereas the FOT line has been fitted by the sublimation model (Eq.(\ref{subl})). 
In b) we have attempted to fit the second peak line by the decoupling theory (Eq.(\ref{dec1})), while in c) a fit of the FOT to the melting theory (Eq.(\ref{Lind})) is also shown.
}
\label{Fig4}
\end{figure}
\newline
$M(T)$ data provide additional information 
about the vortex behavior.  In the reversible region above $T_{irr}$ a 
clear diamagnetic signal is present up to temperatures larger 
than $T_{c}$.  This region is characterized by strong fluctuations and there 
is no well defined upper critical temperature $T_{c2}$. 
The temperature $T_{fluct}$, at which diamagnetic (superconducting) fluctuations appear, has been defined as the temperature where the data begin to deviate from the horizontal normal state line (see Fig.$\ref{fig2}$c). 
The simplest 
way to estimate $T_{c2}$ is to 
use the extrapolation method based on the linear Abrikosov 
formula \cite{Abrikosov}.  The transition temperature $T_{c2}$ is 
derived from the intersection of a linear fit with the normal-state 
horizontal line, as shown in Fig.$\ref{fig2}$c.  It was shown that 
this procedure is not totally correct for HTSC, where the Abrikosov 
linear region is limited to a small temperature range 
because of the rounding close to $T_{c2}$ \cite{HAO, OTT}. 
Indeed in the underdoped regime, where fluctuations are larger, using extrapolation we get unphysical values for the upper critical field (positive slope of $H_{c2}(T)$, see Sec. \ref{discussion}). However, treating the data as proposed by Landau and Ott \cite{OTT} one gets more reasonable upper critical lines for all the samples (see Fig.\ref{Fig4}).  
\newline We also performed isothermic ZFC $M(H)$ measurements at different 
temperatures (see Fig.$\ref{fig3}$).  In the OD samples we could observe two peaks in the $M(H)$ curves (see Fig.$\ref{fig3}$a for OD-31K).
The first minima $H_p$ in the OD samples is known to be related to surface 
\cite{BURLACHKOV} 
and/or geometrical \cite{ZELDOV}
barriers.  Due to these barriers the field doesn't penetrate the 
bulk at the lower critical field $H_{c1}$ but only at an higher field $H_{p}$. 
The second (and largest) minima $H_{sp}$ (second peak) is 
related to some flux-pinning mechanism, although its origin is 
controversial \cite{TAMEGAI,KRUSIN,GIAMARCHI,KOSHELEV}.
In UD samples only one peak could be observed.
We argue that this
is actually the second peak $H_{sp}$. The penetration field $H_p$ is most
probably hidden, due to the low value of $H_{sp}$. This interpretation is supported by the fact
that even in the OD samples it is difficult to identify $H_p$ at high temperatures close
to $T_c$ (where $H_{sp}$ occurs at low fields). Moreover, very accurate SQUID measurements on
UD-29K clearly showed the presence of two minima at $H_p$ and $H_{sp}$ even in the underdoped regime \cite{DIVAKAR}. 
We also performed some full hysteresis loops, as shown in the insets of Fig.$\ref{fig3}$. The
ascending and the descending branches meet at $H_{irr}$, whose values
are consistent with those obtained by FC-ZFC $M(T)$ curves.
\newline In order to facilitate the analysis and discussion of the 
experimental results, the characteristic fields ($H_{c2}(T)$, $H_{fluct}(T)$, $H_{FOT}(T)$ and
$H_{sp}(T)$) of the four samples have been plotted in the \textit{H vs T} phase diagrams shown in 
Fig.$\ref{Fig4}$. 
The magnetic phase diagram of LSCO is usually 
divided in four main phases:
\begin{enumerate}
\item Above the upper critical field 
$H_{c2}(T)$, LSCO is in the non-superconducting state and the 
magnetic flux is free to enter the crystal homogeneously. 
\item Between $H_{c2}(T)$ 
and $H_{FOT}(T)$ ($H_{irr}(T)$) the magnetic flux is partially expelled from the 
superconductor. The magnetic field is present in the sample in the 
form of vortices which are in a reversible regime.  In 
this region the vortices are thermally activated and highly dynamic.  
\item  Below $H_{FOT}(T)$ ($H_{irr}(T)$) the vortices are in an 
irreversible regime, as can be seen by the difference in the FC/ZFC 
data or in the hysteresis loops.  Here the vortices are frozen in a 
lattice (VL), which can be directly observed in SANS experiments 
\cite{GILARDI1,GILARDI2,GILARDI3}.  
\item Below $H_p(T)$ (ideally $H_{c1}(T)$) the system is in the Meissner state and the flux is
completely expelled from the bulk of the sample.
\end{enumerate}
Indeed we can roughly understand our results in LSCO within this description, even
though we have some additional lines in the phase diagram (e.g. $H_{sp}$ and $H_{fluct}$).
The first observation is that the magnetic phase diagrams of OD
and UD LSCO are \textit{qualitatively} similar but 
\textit{quantitatively} very different.  In particular for the UD samples the reversible   
region is much larger than for OD ones, whereas the second peak line occurs at much lower fields.

\section {Discussion}
\label{discussion}

Before discussing the possible reasons for this strong doping dependence of 
the phase diagram we want to have a detailed look at the single lines. 
\newline
We start from the upper critical line $H_{c2}(T)$, which is not well 
defined since fluctuations are very strong near $T_{c2}$. This is more pronounced in the underdoped regime, where
diamagnetic fluctuations are present even 
at temperatures $T_{fluct}$ much larger than $T_{c}$. This anomalous behavior in the underdoped regime has also 
been observed in Nernst  \cite{XU,WANG,WEN} and scanning SQUID microscopy \cite{IGUCHI} experiments and has been interpreted as being due to vortex-like excitations in the pseudogap region.
As a consequence, $H_{c2}(T)$ as determined by extrapolation has an unphysical positive slope.
In order to get more reasonable upper critical field lines, we used the Landau-Ott scaling procedure for magnetization data \cite{OTT}, taking the values of $H_{c2}$(0~K) listed in Table \ref{table}. The resulting $H_{c2}(T)$ lines are plotted in Fig.$\ref{Fig4}$.
\newline
We turn now to the FOT line $H_{FOT}(T)$, which is 
usually identified with the melting line \cite{BRANDT,HOUGHTON,BLATTER}, 
that is the transition of the vortex-solid into a vortex-liquid, in which 
the VL loses its shear modulus. 
 The temperature dependence of  $H_{FOT}(T)$ is predicted by the melting theory to be \cite{BRANDT,HOUGHTON,BLATTER}:
\begin{equation}
    H_{melt}(T)=H_{m}\cdot \biggl(1-\frac{T}{T_{c}}\biggr)^{m}
   \label{melting}
\end{equation}
The prefactor is known to 
depend almost only on the anisotropy of the system. In fact, 
considering $H_{m}\sim \gamma^{-2}T_{c}^{-2}\lambda_{ab}^{-4}$ ($\lambda_{ab}$ is the in-plane penetration depth) and 
the fact that $T_{c}^{-2}\lambda_{ab}^{-4}$ is almost constant 
\cite{comment}, one obtains $H_{m}\sim \gamma^{-2}$. Fitting our data by this model 
is not satisfactory, since we obtain a huge doping dependence of the 
exponent $m$  and the prefactor $H_{m}$ 
doesn't follow the expected $\gamma$ dependence (see Table \ref{table}).
Moreover, in all SANS experiments on HTSC 
\cite{CUBITT,AEGERTER,GILARDI2} the ring-like intensity expected between $H_{FOT}$ and $H_{c2}$ for a liquid of \textit{straight} 
vortices \cite{NORDBORG} has never been observed.
A more precise melting theory, still based on the Lindemann criterion \cite{LINDEMANN}, predicts a more complicated temperature dependence of the melting line \cite{BLATTER}: 
\begin{equation}
    H_{melt}(T)=\frac{4c_L^4H_{c2}(0)\frac{B}{G}(\frac{T_c}{T}-1)^2}{\biggl(1+\sqrt{1+4c_L^4\frac{B}{G}(\frac{T_c}{T}-1)\frac{T_c}{T}}\biggr)^2}
   \label{Lind}
\end{equation}
where $G=\frac{1}{2}(\frac{\gamma k_B T_c}{(4\pi / \mu _0)H_c^2(0)\xi_{ab}^3(0)})^2$ is the Ginzburg number ($\mu_0$ is the permeability of free space, $k_B$ is the Boltzmann's constant, $H_c$ is the thermodynamic critical field, and $\xi_{ab}$ is the in-plane coherence length), B$\approx$5.6 and $c_L$ is the Lindemann number.
However, even this formula doesn't describe our data very well, since the fitted curves are unsatisfactory (see for example Fig.$\ref{Fig4}$c for UD-29K), $c_L$ is doping dependent and in some cases higher than the expected values ($c_L\sim$0.1-0.2).
\begin{table}[t]
    \centering
    \caption{Characteristic parameters for LSCO as a function of the 
    Sr concentration $x$. The values of the upper critical field $H_{c2}$(0~K)~\cite{ANDO}, of the penetration depth 
    $\lambda_{ab}$~\cite{Panagopoulos}, and of the anisotropy $\gamma$~\cite{SASAGAWA2,KIMURA,Willemin} have been 
    extrapolated from experimental values found in the literature. 
    $H_m$ and $m$ have been obtained by fitting the data by Eq.(\ref{melting}), the Lindemann number $c_L$ using Eq.(\ref{Lind}).
$\gamma_{subl}$ and $\gamma_{dec}$ 
    are the anisotropies 
    obtained by fitting our data using the sublimation, respectively 
    decoupling models. Finally, the exponent $n$ has been obtained by fitting the data with Eq.(\ref{power}).
}   
\vspace{3mm}
\begin{ruledtabular}
\begin{tabular}{c|cccc}
LSCO &OD-31K&OD-36K&UD-29K&UD-19K\\
\hline
$x$ & 0.20&0.17 &0.10 &0.075 \\
$T_{c}$  & 31.5 K& 36.2 K& 29.2 K& 19 K\\
$\Delta T_{c}$  & 2.8 K& 1.5 K& 1.3 K& 3.8 K\\
$H_{c2}$(0 K) & $\sim$ 45 T& $\sim$ 60 T & $\sim$ 45 T & $\sim$ 35 T \\
$\lambda_{ab}$ &1970 \AA &$\sim$ 2400 \AA & 2800 \AA& $\sim$ 3000 \AA\\
$\gamma$ & 20(2)	& 20(2) & 45(5) & 60(5) \\
$H_m$ & 30 T& 28 T&  15 T&15 T \\
$m$ & 1.7& 1.8&3.3 &6.1 \\
 $c_{L}$ & 0.28 &0.29 &0.20 &0.16 \\
$\gamma_{subl}$ & 20&22 &47 &64 \\
$\gamma_{dec}$ & 12 &13 &40 &85 \\
$n$ & 2.3&2.1 &2.1 &2.5 \\
\end{tabular}
\end{ruledtabular}
 \label{table}
\end{table}
\begin{figure}
\includegraphics[width=3.5in]{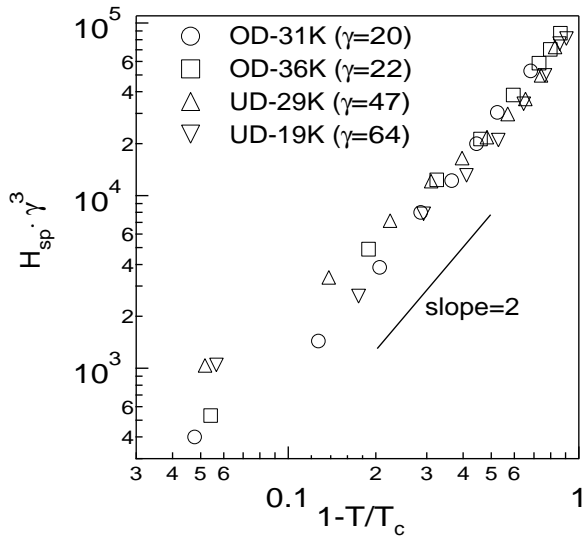}
\caption{Temperature dependence of $H_{sp}\cdot \gamma^{3}$ plotted in a double logarithmic scale. All the data measured in samples with different doping levels collapse on one line with
slope~$\sim$~2. This indicates that the power law (Eq.(\ref{power})) has an exponent $n\approx$~2 and $H_0\propto\gamma^{-3}$.
We have used the values of $\gamma$
obtained by fitting our data using Eq.(\ref{subl}) ($\gamma_{subl}$ in Table \ref{table}).}
\label{fig5}
\end{figure}
\newline
An alternative model to the melting transition is given by the sublimation theory 
\cite{SASAGAWA1,SASAGAWA2}, based on the strong anisotropy originating from the 
layered structure intrinsic to all HTSC. Within this scenario the melting 
is accompanied by the simultaneous decoupling of the vortex lines into 
2D pancake vortices (\textit{vortex gas}). The phenomenological scaling law which applies to 
all HTSC is given by:
\begin{equation}
H_{subl}(T)[Oe]=2.85\cdot \gamma^{-2}s^{-1}\biggl(\frac{T_c}{T}-1\biggr)
 \label{subl}
\end{equation}
where $s$ is the distance between the $CuO_2$ layers (6.6$\cdot10^{-8}$ cm in LSCO). This formula has been used in order to explain the FOT 
transition and nicely fits our data. $\gamma$ is the only free parameter, and the fitted values are in good agreement with the measured values of the anisotropy (see Table \ref{table}).
\newline 
It remains to discuss the second peak line $H_{sp}(T)$ which has been explained 
on the basis of the thermal decoupling theory \cite{Glazman,Daemen,Sutjahja}, which predicts the 
suppression of long-range order in the direction of the applied 
field due to thermal fluctuations.  The expected temperature 
dependence is \cite{Sutjahja}
\begin{equation}
    H_{dec}(T)=H^{*}\cdot \biggl(\frac{T_{c}}{T}-1\biggr)
   \label{dec1}
\end{equation}
with 
$H^{*}=\Phi_{0}^{3}/(16\pi^{3}ek_{B}\mu_{0}s\gamma^{2}T_{c}\lambda_{ab}(0)^{2})
   \label{dec2}$, where $\Phi_{0}$
is the flux quantum and $e\approx$~2.718 is the exponential number.
This function doesn't fit well our data, as shown in Fig.$\ref{Fig4}$b
for OD-36K. Moreover, the estimated values for $\gamma$, obtained by 
substituting the known values of $s$, $T_{c}$ and $\lambda_{ab}(0)$ in the 
theoretical expression for $B^{*}$, are not satisfactory compared 
to the experimental values (see 
Table \ref{table}).
Moreover recent SANS 
measurements \cite{GILARDI2} indicate that the diffraction signal from the vortex lattice persists up to 
$H_{FOT}(T)$ and therefore discredit the decoupling theory.
The origin of the second peak is most probably to be found in some change 
of the pinning mechanism. 
It has been often suggested that this feature is related to the transition to a more disordered \textit{vortex glass} phase \cite{GIAMARCHI,KOSHELEV}, and very recent experimental results confirm this interpretation \cite{GILLER,DIVAKAR}.
Our experimental data are better fitted by a power law \cite{KODAMA}
\begin{equation}
    H_{sp}(T)=H_{0}\cdot \biggl(1-\frac{T}{T_{c}}\biggr)^{n}
   \label{power}
\end{equation}
as can be seen in Fig.$\ref{Fig4}$ and Fig.$\ref{fig5}$. 
The value of the exponent is close to $n$=2 in all samples 
(see Fig.$\ref{fig5}$ and Table \ref{table}). 
Interestingly, the value of $H_{0}$ seems to be proportional to 
$\gamma^{-3}$, even though (up to our knowledge) no theory predicts such a $\gamma$ dependence. However, a large anisotropy naturally renders the vortex system more susceptible to disorder. The observed anisotropy dependence of $H_{sp}(T)$ is therefore in qualitative agreement with a scenario where  the second peak line is related to a field-induced vortex glass transition.

\section{Conclusion}

A first look at the magnetic phase diagrams shown in 
Fig.$\ref{Fig4}$ could indicate that the vortex matter in LSCO is 
strongly doping dependent. This is true from a quantitative point of 
view, but qualitatively all samples display the same transitions 
(second peak, irreversibility, FOT and upper critical lines).
The quantitative doping dependence of the magnetic phase diagram can mainly be explained by 
the different degree of anisotropy: $H_{sp}$ is found to be proportional 
to $\gamma^{-3}$ and $H_{FOT}$ to $\gamma^{-2}$.
The interpretation of the second peak in LSCO is still controversial 
but our data seem to favor the vortex glass scenario, whereas the 
FOT line is consistent to the sublimation theory rather than to the melting theory.
Moreover, strong superconducting fluctuations above $T_c$ have been observe in the underdoped regime.

\vspace{5mm}

\section{Acknowledgments}
We would like to thank C.D. Dewhurst for valuable discussions.
This work was performed at the Paul Scherrer
Institute, Switzerland, and was supported by the Swiss National Science
Foundation and the Ministry of Education, Technology
and Science of Japan (NM, MO).

\end{document}